%% file: paper.tex
\pdfoutput=1
\documentclass[aps,prd,amsmath,floats,floatfix, twocolumn,
superscriptaddress,nofootinbib,showpacs,longbibliography]{revtex4-1}

\usepackage[T1]{fontenc}
\usepackage[utf8]{inputenc}
\usepackage{lmodern}

\usepackage{verbatim}
\usepackage{makecell}

\usepackage[dvipsnames, usenames]{xcolor}
\definecolor{linkcolor}{rgb}{0.0,0.3,0.5}
\usepackage[hypertexnames=false, unicode, colorlinks=true, linkcolor=linkcolor,
citecolor=linkcolor, filecolor=linkcolor,urlcolor=linkcolor,
pdfusetitle]{hyperref}

\usepackage[all]{hypcap}
\usepackage{graphicx}
\usepackage{xspace}
\usepackage{braket}
\usepackage{amssymb}
\usepackage{bm} 

\graphicspath{{Figures/}}

\usepackage{microtype}

\usepackage{blindtext}
\usepackage{mathrsfs}
\usepackage{orcidlink}

\usepackage{tikz}
\usetikzlibrary{positioning}
\tikzset{%
  neuron/.style={
    circle,
    draw,
    minimum size=1cm
  },
  neuron missing/.style={
    draw=none,
    scale=4,
    text height=0.333cm,
    execute at begin node=\color{black}$\vdots$
  },
}

\DeclareMathAlphabet{\mathpzc}{OT1}{pzc}{m}{it}

\newcommand{\vdot}{\dot{v}}

\newcommand{\lie}{\mathcal{L}}

\newcommand{\NRSurOld}{\texttt{NRHybSur3dq8}\xspace}

\begin{document}

\title{Learning post-Newtonian corrections from numerical relativity}

 \newcommand{\Cornell}{\affiliation{Cornell Center for Astrophysics and Planetary
     Science, Cornell University, Ithaca, New York 14853, USA}}

\author{Jooheon Yoo \orcidlink{0000-0002-3251-0924}} \email{jy884@cornell.edu} \Cornell
\author{Michael Boyle \orcidlink{0000-0002-5075-5116}}\Cornell
\author{Nils Deppe \orcidlink{0000-0003-4557-4115}} \Cornell

\hypersetup{pdfauthor={Yoo et al.}}
\date{\today}

\begin{abstract}
Accurate modeling of gravitational waveforms from compact binary coalescences
remains central to gravitational-wave (GW) astronomy. Post-Newtonian (PN)
approximations capture the early inspiral dynamics analytically but break down
near merger, while numerical relativity (NR)
provides the accurate yet computationally expensive waveforms over limited
parameter ranges. We develop a physics-informed neural network (PINN) framework
that learns corrections mapping PN dynamics and waveforms to their NR counterparts.
As a demonstration of the approach, we use the TaylorT4 PN model as the baseline, and train the network on
a remarkably small dataset of only eight hybridized NR
surrogate waveforms (NRHybSur3dq8) to learn higher-order corrections to the orbital dynamics
and waveform modes for nonspinning noneccentric systems. Physically motivated loss terms enforce known limits and symmetries, such as vanishing
corrections in the Newtonian limit and suppression of odd-$m$ modes in equal-mass systems, promoting consistent and reliable
extrapolation beyond the training region.
We simultaneously incorporate corrections that account for the different meaning of mass parameters in PN and NR descriptions.
The learned corrections
significantly reduce the phase and amplitude error through the inspiral
up to about $200M$ before the merger. This approach provides a differentiable and computationally
efficient bridge between PN and NR, offering a path toward waveform models that
generalize more robustly beyond existing NR datasets.
\end{abstract}

\maketitle

\section{Introduction}
\label{sec:introduction}

Since the landmark discovery of gravitational waves (GWs) from binary black holes (BBHs)
in 2015~\cite{Abbott:2016blz}, detections have become increasingly frequent as the sensitivity
of ground-based interferometers continues to improve. Recent publication of the loudest
observed event~\cite{LIGOScientific:2025obp} highlights the remarkable progress achieved
over the past decade. With next-generation detectors on the horizon~\cite{LISA:2017pwj,Reitze:2019iox,
Punturo:2010zz}, both the detection rate and sensitivity are expected to increase further.
Thus, the demand for high-fidelity GW templates has become more important than ever.

Numerical relativity (NR) is the only \textit{ab initio} method for solving Einstein's
equations for the mergers of two compact objects and has played a fundamental role in
GW theory and GW astronomy~\cite{Pretorius:2005gq, Campanelli:2005dd,Baker:2005vv,
SXSCatalog,Scheel:2025jct}. Despite many efforts to make these simulations more
computationally efficient, they remain prohibitively expensive for direct applications
such as source parameter estimation. To address this bottleneck, several waveform models
have been developed that provide a faster alternative to evolving an entire NR
simulation to generate a GW template.

Effective-one-body (EOB)~\cite{Buonanno:1998gg, Buonanno:2000ef,
Buonanno:2005xu, Damour:2000we, Damour:2001tu, Buonanno:2014aza,
Damour:2008yg, Ramos-Buades:2023ehm} and phenomenological
models~\cite{Khan:2015jqa, Husa:2015iqa, Hannam:2013oca,
Pratten:2020ceb, Pratten:2020fqn, Ajith:2007qp, Ajith:2007kx,
Ajith:2009bn, Santamaria:2010yb, London:2017bcn, Khan:2018fmp,
Khan:2019kot, Dietrich:2018nrt, Dietrich:2019nrt, Thompson:2020nei,
Garcia-Quiros:2020qpx, Garcia-Quiros:2020qlt} incorporate some known
physics, but still require ``calibration'' to NR waveforms, including
careful selection of the parameters to be calibrated, with limited
ability to capture physics beyond the currently known analytical
behavior.
NR surrogate models~\cite{Blackman:2015pia, Varma:2018mmi, Blackman:2017dfb, Blackman:2017pcm,
Varma:2019csw} are a more recent
addition to the family of BBH waveform models
and take a data-driven approach by training directly on the waveform
outputs from NR simulations. Consequently, these surrogates reproduce NR waveforms
more faithfully than other semianalytical models; however, they are limited to regions of
parameter space where NR simulations exist. In addition, as NR simulations typically cover
only the final $\sim20$ orbits before merger, NR-only surrogate models cannot
span the full LIGO frequency band. This limitation will become increasingly problematic since
both future space-based and ground-based detectors promise sensitivity at even lower frequencies.
To address this, hybrid surrogate models have been developed in which hybrid waveforms
combining PN-based early inspiral and NR merger-ringdown data are used for training.

In this sense, all three classes of models---surrogate, phenomenological, and EOB---are hybrid models, incorporating some analytical information, and some numerical information.
However, discrepancies between the analytical and numerical descriptions can lead to imperfect
stitching, introducing hybridization errors that can compromise the accuracy of
the resulting models~\cite{Sun:2024kmv}.
Even the parameters used in the PN and NR descriptions differ;
Ref.~\cite{SunStein2025} found that the mass parameters in the two
formalisms can differ even at 1PN order, to say nothing of differences
in spins and eccentricities.  These discrepancies lead to systematic
errors when not taken into account in comparing analytic and numerical
waveforms.  Essentially, the part of the model comprised of analytical
information and the part comprised of numerical information describe
physically different systems.

\begin{figure*}[!t]
\label{fig:highest_q_validation_case}
\centering
 \includegraphics[width=1.0\linewidth]{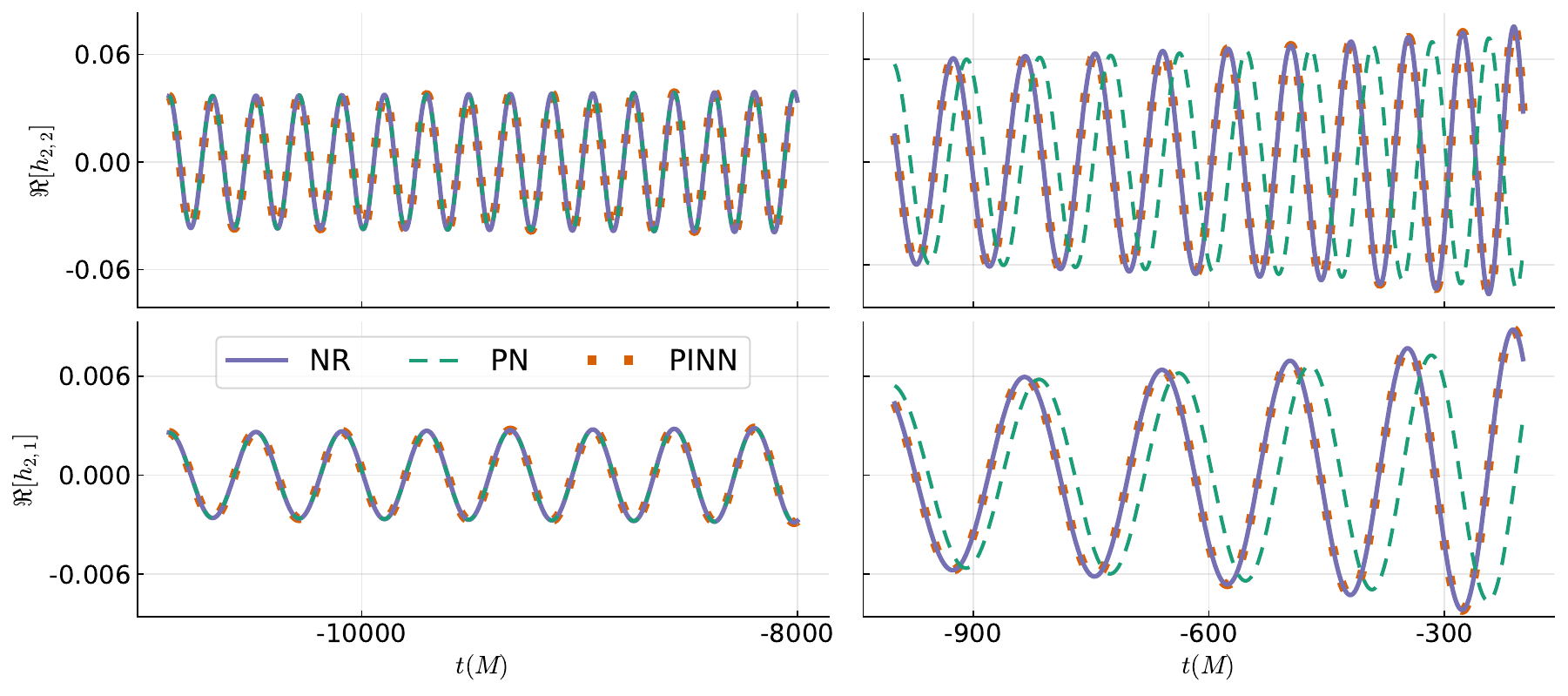}
\caption{Real parts of $h_{2,2}$ and $h_{2,1}$ for the highest mass ratio validation case
of $q=7.93$ for the final NN model of the PN-to-NR correction.
The waveform mismatch is reduced from $2\times 10^{-1}$ to
$1\times 10^{-6}$. All waveforms are phase-aligned
at the start. Although the PN waveform agrees well with NR
in the early inspiral, it deviates significantly near merger
where both phase and amplitude discrepancies emerge. The NN corrections significantly
reduce these discrepancies, allowing the PN waveforms to remain
consistent with NR counterparts in the late inspiral regime.}
\end{figure*}

As pointed out in recent work~\cite{Sun:2024kmv}, there is a trade-off in selecting the hybridization
windows. Choosing an earlier window increases errors due to nonzero residual eccentricity in
NR simulations, whereas a later window amplifies discrepancies arising from PN's inaccuracy near merger. This naturally raises the question: can we introduce data-driven higher-order corrections to
PN approximations so that they more closely match NR waveforms in the late
inspiral regime?

In recent years, neural networks have emerged as versatile tools in computational physics,
capable of modeling complex nonlinear dependencies. In the field of GW astronomy, they have
been applied to a wide range of problems including waveform modeling, parameter 
estimation~\cite{Thomas:2025rje, Negri:2025cyc}.
One application is the use of physics-informed neural networks
(PINNs), which incorporate physical results or principles into the
neural network design or training process~\cite{ghosh2020, coelho2025,
marion2023, rackauckas2021, bolibar2023, osti_1991293, rodrigues2025,
kapoor2023}.\footnote{Neural ODEs, universal differential equations,
and physics-informed neural networks are sometimes distinguished in
the literature; here we simply refer to them collectively as PINNs.}
In particular, Keith \textit{et al.}~\cite{Keith:2021arn} explored the use of neural networks to learn
relativistic orbital corrections to Newtonian dynamics directly from
GW data. Building upon this idea, we investigate whether neural networks can learn
higher-order corrections to PN approximations to bridge the gap between PN and NR waveforms.

To explore this potential, we develop a framework based on neural networks that learns the
physics-informed corrections that map PN dynamics and waveforms to their NR counterparts.
The network is designed to augment the existing PN formalism by learning residual
correction terms that can account for the higher-order missing physics in the standard
PN expansion.

By anchoring the model to known PN dynamics, we effectively embed physical information
into the learning process, ensuring that the network's corrections remain consistent
with well-understood analytical behavior. In addition, physics-based loss terms impose
constraints on these corrections, which regularize the solution and promote physically
meaningful extrapolation beyond the training region. As a first step forward in this framework,
we focus on nonspinning, noneccentric systems. In this work, we train
the model on a small set of eight hybrid NR surrogate waveforms, demonstrating
the framework's potential for data-efficient PN-to-NR calibration.
The resulting improvement in waveform agreement is illustrated in Fig.~\ref{fig:highest_q_validation_case},
which shows the highest mass ratio validation case. In this example, the mismatch is reduced
from $2\times 10^{-1}$ for the baseline PN model to $1\times 10^{-6}$ after applying
the learned neural network corrections.

The rest of the paper is organized as follows.
In Sec.~\ref{sec:2pn3pn}, we describe results from a toy problem where we
  learn the 3PN terms using a neural network-augmented 2PN evolution.
  This serves as a controlled test of our framework and demonstrates the network’s
ability to reproduce analytically known higher-order corrections.
Section~\ref{sec:pntonr} presents the full
framework that extends this approach to learning corrections from the PN
model to NR waveforms, including network architecture, data preparation,
and implementation details. Its subsections
describe the loss functions, examine the framework's extrapolation
behavior beyond the training region, and explore how incorporating
a small amount of additional training data can further improve model accuracy.
Finally, in Sec.~\ref{sec:conclusion}, we summarize our findings
and discuss potential avenues for extending the framework to a broader
region of parameter space.

\section{2PN \texorpdfstring{$\rightarrow$}{to} 3PN Experiment}
\label{sec:2pn3pn}

The primary objective of this study is to evaluate whether a
neural network can effectively learn physics-informed corrections that reduce
discrepancies between PN and NR waveforms near merger.

Before applying the framework to NR data, we first test it on a simpler,
analytically controlled problem: learning the corrections required to map the 2PN approximation to
the 3PN approximation. This serves as a proof of concept, verifying that the network can
recover known higher-order terms when trained on data generated purely from PN theory.
Since both 2PN and 3PN expressions are analytically known, this experiment provides a clean
test bed for assessing the network's capacity to represent higher-order corrections.

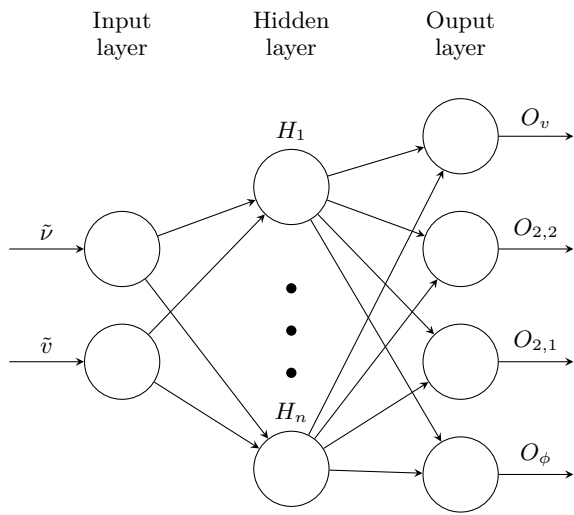
\begin{figure}[h!]
\label{fig:2pn3pn_design}
\centering\centering
\input{fig_2pn3pn_tikz.tex}
\caption{Neural network architecture for the 2PN$\rightarrow$3PN experiment.
The two inputs, $(\nu, v)$, are linearly rescaled to $(\tilde{\nu},\tilde{v})$ to the interval
$[-1, 1]$ to improve numerical stability during training. Each network output is multiplied by a tunable
scaling coefficient before being applied as a correction to either the orbital dynamics
or the waveform modes. The final architecture uses a single hidden layer with eight neurons.}
\end{figure}

The difference between the 2PN and 3PN waveforms arises from two main sources:
(i) missing higher-order terms in the orbital dynamics, represented by the right-hand side of
the evolution equation for the orbital velocity $\vdot$, and
(ii) missing higher-order terms in the waveform amplitude and phase.
To address both effects simultaneously, we employ a neural-network architecture with
multiple outputs: one branch predicting corrections to the orbital dynamics, and another
predicting corrections to the dominant waveform modes, $h_{2,2}$ and $h_{2,1}$.
The overall design of this 2PN $\rightarrow$ 3PN experiment
 is illustrated in Fig.~\ref{fig:2pn3pn_design}.

To ensure stable training and well-conditioned gradients, both inputs, the
symmetric mass ratio $\nu$ and the orbital velocity $v$, are linearly rescaled
to the interval $[-1,1]$ before being passed into the network.
The rescaled inputs, denoted by $\tilde{\nu}$ and $\tilde{v}$ are obtained by
applying linear transformation such that $\nu \in [0.09, 0.25]$ and $v \in [0.24, 0.36]$
map to the range $[-1,1]$. These bounds are chosen to be slightly wider than
the actual training data range to avoid boundary effects
and to maintain consistent behavior near the edges.

The orbital dynamics are governed by a PN expansion in powers of $(v/c)$,
truncated at a given PN order:
\begin{equation}
\label{eqn:vdot}
\vdot = \frac{32c^4\nu}{5GM} \bigg(\frac{v}{c}\bigg)^9 \bigg[1 + c_2 (v/c)^2 + \ldots + c_{2n} (v/c)^{2n} \bigg],
\end{equation}
where the prefactor is the leading-order Newtonian contribution, $n$
the PN order, and $M$ the total mass of the system.
The neural network is trained to learn the missing contribution between the
2PN and 3PN truncations, $O_v(\nu,v)$.
The NN correction to the orbital dynamics can be written as
\begin{equation}
\label{eqn:vdot_corr}
\vdot_{\text{corr}} = \vdot_{2\text{PN}} + \frac{32c^4\nu}{5GM} \bigg(\frac{v}{c}\bigg)^9\,\bigg[\beta_v \,O_{v}(\nu,v)\,(v/c)^5\bigg],
\end{equation}
where $\beta$ is a tunable scaling coefficient, and $O_v(\nu,v)$ is
the neural-network prediction representing the higher-order correction.

In earlier attempts, we allowed the network to output two independent
coefficients corresponding to the  $(v/c)^5$ and $(v/c)^6$ terms separately.
However, we found that these corrections were highly degenerate:
the network tended to fit combinations of coefficients that produce nearly identical
effects on the orbital evolution. Empirically, we observed that one term
is sufficient to model the overall difference between $\vdot_{2\text{PN}}$ and
$\vdot_{3\text{PN}}$.

The PN waveform modes are likewise expanded and truncated in $(v/c)$,
necessitating corrections to both amplitude and phase for each mode.
The neural-network-predicted corrections are applied as
\begin{align}
\label{eqn:waveform_correction}
\begin{aligned}
h'_{2,2} &= h_{2,2}\big(1+\beta_{2,2} O_{2,2}(\nu,v)\big)e^{2\beta_{\phi} O_{\phi}(\nu,v)} \\
h'_{2,1} &= h_{2,1}\big(1+\beta_{2,1} O_{2,1}(\nu,v)\big)e^{\beta_{\phi} O_{\phi}(\nu,v)}
\end{aligned}
\end{align}
Each of the neural-network outputs $O_{2,2}$, $O_{2,1}$, and
$O_\phi$ is multiplied by a respective scaling coefficient, $\beta_{\ell,m}$ or $\beta_{\phi}$,
before being applied.
These coefficients act as tunable hyperparameters that control the relative
strength of each correction during training, ensuring balanced convergence
between the amplitude and phase components.

The overall 2PN$\rightarrow$3PN procedure can be summarized as:
\begin{enumerate}
\item{Apply the correction $O_v(\nu,v)$ to the 2PN expression for $\vdot$ following
Eq.~\eqref{eqn:vdot_corr}.}
\item{Integrate the resulting ODE to obtain the corrected orbital evolution.}
\item{Compute the waveform modes $h_{2,2}$ and $h_{2,1}$ from the corrected orbital evolution.}
\item{Apply the waveform mode correction as described in Eq.~\eqref{eqn:waveform_correction}.}
\end{enumerate}

\begin{figure*}[!t]
\label{fig:worst_case_PN}
\centering
 \includegraphics[width=1.0\linewidth]{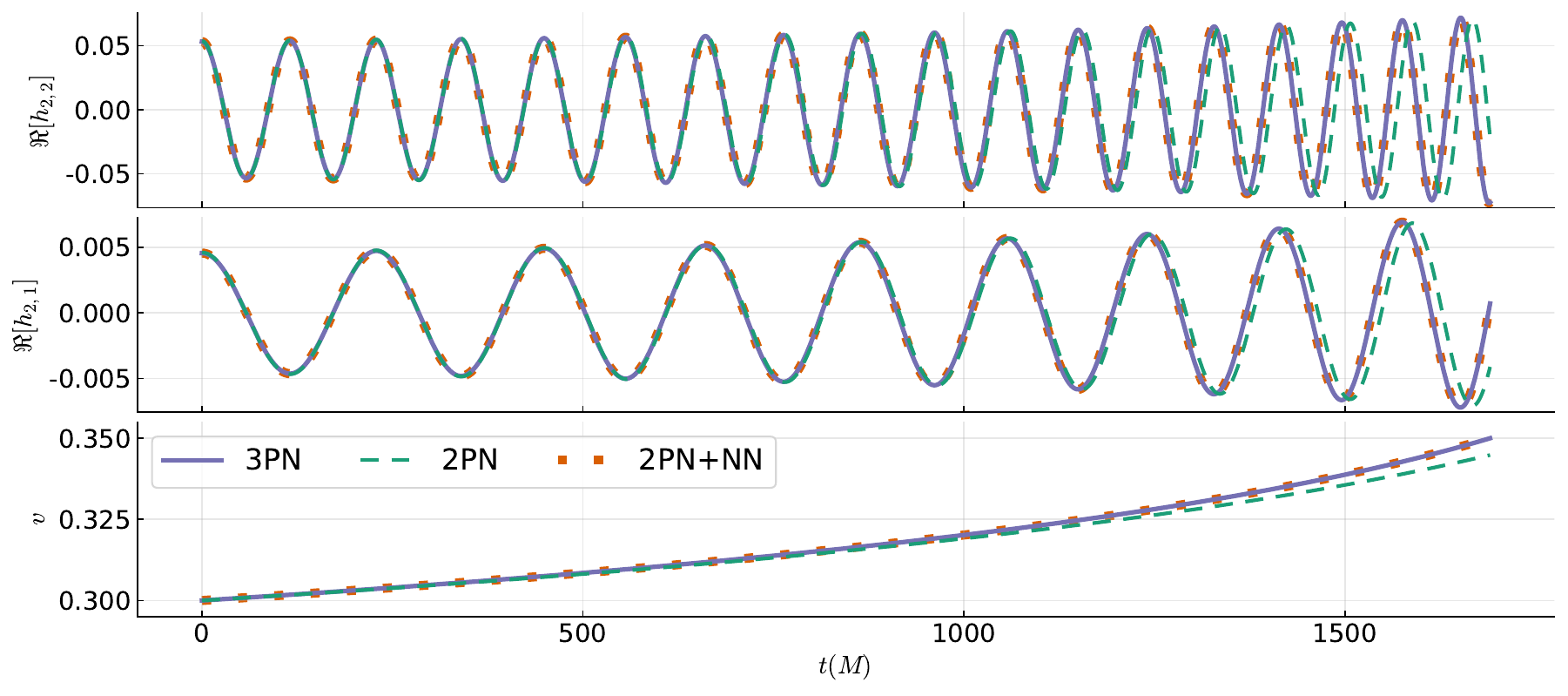}
\caption{Real parts of $h_{2,2}$ and $h_{2,1}$, and the
orbital velocity $v$, for the worst
validation case ($q=7.84$). Including the NN corrections
reduces the mismatch from $10^{-1}$ to $4\times10^{-8}$.
All the waveforms are phase-aligned at $t=0$. The consistent
improvement in the corrected orbital velocity shows that
the NN captures genuine orbital corrections rather than
simply compensating through the waveform-mode corrections.}
\end{figure*}

The training data consist of 3PN waveforms sampled on a uniform grid in the
symmetric mass ratio $\nu$. Specifically, we generate
waveforms for values of $q \in [1,8]$ uniformly distributed in $\nu$
and segment each waveform into uniform $\Delta v$ intervals
within $v\in[0.3,\,0.35]$. The network is first
trained on these segmented waveforms to accelerate the learning of the
$v$-dependence of the corrections, following a strategy conceptually
similar to multiple-shooting methods. The network is then refined
using full waveforms to ensure global consistency in the corrections
across the inspiral.

The network is trained by minimizing a loss function
that quantifies the discrepancy between the neural-network-corrected
2PN waveform and the reference waveform. The primary contribution
to the loss is the mean waveform mismatch over all training samples,
supplemented by a  regularization term to mitigate overfitting.

For each sample, we compute the mismatch between the reference 3PN waveform and the
neural-network-corrected 2PN waveform obtained through the correction procedure
outlined above. The mismatch between two waveforms is computed as
\begin{equation}
\label{eqn:waveform_error}
\Delta(h_1,h_2) = \frac{1}{2} \frac{\lVert h_1 - h_2
  \rVert^2}{\sqrt{\lVert h_1 \rVert^2  \lVert h_2 \rVert^2 }}\,.
\end{equation}
where $\lVert \, \rVert$ denotes the complex inner product
\begin{equation}
\label{eqn:l2}
\lVert h \rVert^2 =   \sum_{t=t_0}^{t_f} \sum_{l,m} |h_{l,m}(t)|^2.
\end{equation}
The waveform loss, $\lie_\text{w}$, is then defined as the average mismatch over all training samples,
\begin{equation}
\lie_{\text{w}} = \langle \Delta (h_1, h_2) \rangle
\end{equation}

Minimizing the waveform mismatch alone can lead to overfitting.
To mitigate this, we include an $L_2$
regularization term, a standard technique in machine learning
that penalizes large neural-network weights.
The total loss function is therefore
\begin{equation}
\label{eqn:2pn3pn_loss}
\lie_{\text{total}} =
\lie_{\text{w}}  +
\lambda_{L_2} \lie_{L_2},
\end{equation}
where
\begin{equation}
\lie_{L_2} =
\sum_n \big(\theta^{\mathrm{NN}}_n\big)^2
\end{equation}
is the squared $L_2$ norm of the neural-network weights, $\theta^{\mathrm{NN}}_n$.
The coefficient $\lambda_{L_2}$ is a hyperparameter that controls the strength of the
regularization. Because regularization aims to reduce sensitivity to inputs,
the penalty is applied only to the weights and not to the biases.

\begin{figure}[h!]
\label{fig:2pn3pn_loss_history}
\includegraphics[width=0.5\textwidth]{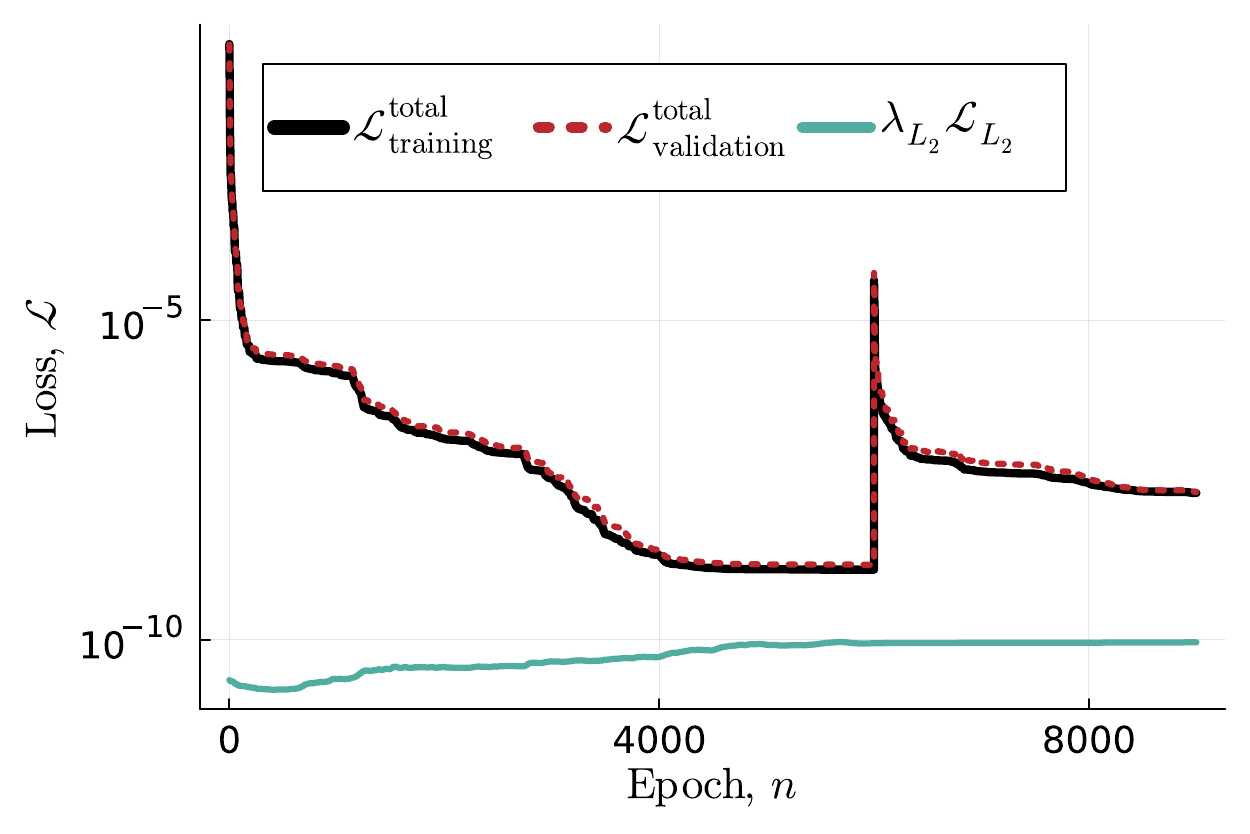}
\caption{Training and validation loss history for the 2PN$\rightarrow$3PN experiment.
The validation losses closely follow the training losses, indicating that the neural network
generalizes beyond the training set. The sharp increase in loss near epoch, $n=6000$,
corresponds to the transition from segmented-waveform training to full-waveform training.}
\end{figure}

We use the quasi-Newton BFGS routine to minimize the total loss function, $\lie_\mathrm{total}$.
We also experimented with adaptive methods such as the
Adam~\cite{kingma2014adam} optimizer, but found that, given
the small training data size and relatively small network size,
the minibatching advantages of Adam could not be fully exploited for the
given framework.

The network achieves an average waveform mismatch on the order of $10^{-8}$ between
3PN and NN-corrected 2PN waveforms over the training range of $q\in[1,8]$.
To confirm generalization,
$20\%$ of the data are withheld for validation at the start of the training.
A representative worst validation
case, including the waveform modes and orbital velocity is
shown in Fig.~\ref{fig:worst_case_PN}.
The training and validation losses converge to the same tolerance
as shown in Fig.~\ref{fig:2pn3pn_loss_history}.
Here, an epoch refers to one complete pass through the full
training dataset; since our implementation does not use minibatching,
each epoch corresponds to a single update (i.e. one optimization step).
The sharp increase in loss around epoch
$n=6000$, marks the transition
from segmented-waveform training to full-waveform training.

\begin{figure}[!h]
\label{fig:NN_orbitalcorrection}
\centering
 \includegraphics[width=1.0\linewidth]{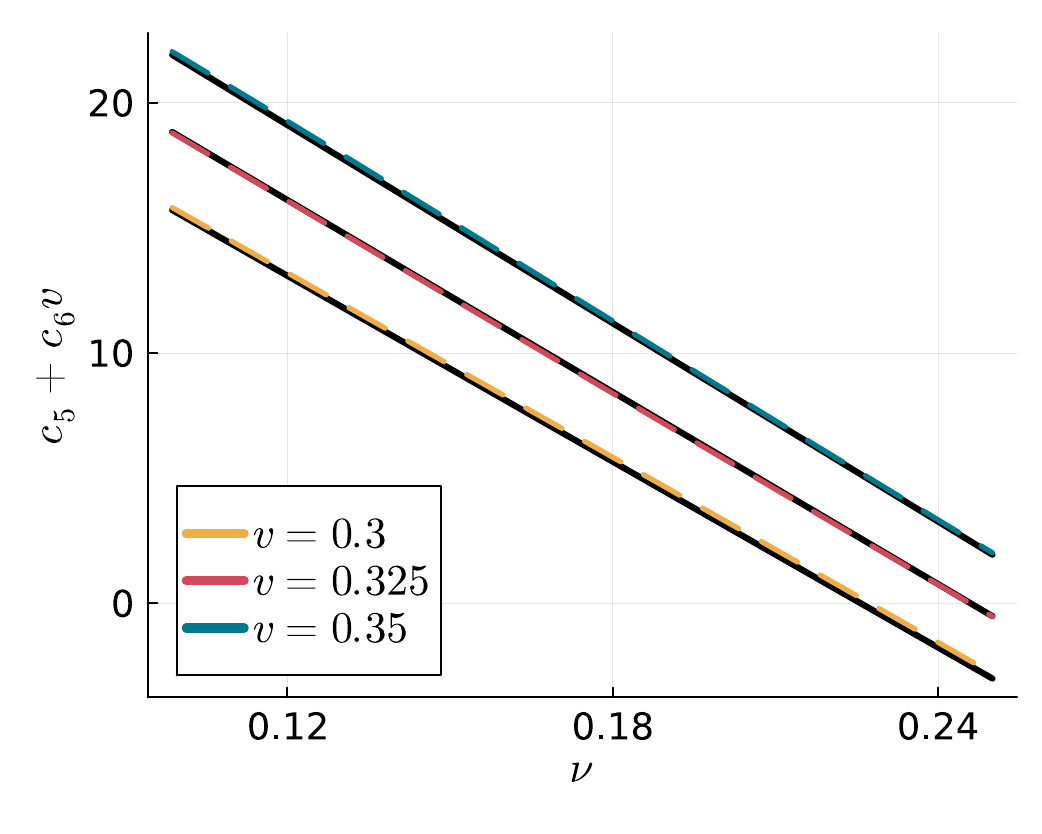}
\caption{Comparison of the true and neural-network–learned coefficients for the
  missing term in the $\vdot$ expansion. Solid lines represent the
  true value while dashed lines the learned coefficients. Different colors
  corresponding to different values of $v$. The close agreement between
  the true and learned curves across different values of $v$ demonstrates
  that the network successfully recovers the missing higher-order orbital corrections
  despite the degeneracy between orbital and waveform mode corrections.}
\end{figure}

Because the loss function is defined solely in terms of waveform mismatch,
the orbital and mode corrections can exhibit degeneracy, where different
combinations of corrections yield nearly identical waveform errors.
For example, the NN could overemphasize waveform mode correction
to compensate for an inaccurately learned orbital correction.
To verify that this degeneracy did not lead to poor modeling of the orbital
dynamics, we perform additional post-training diagnostics. In particular, we
compare the corrected $v$-trajectories with the reference 3PN results in
Fig.~\ref{fig:worst_case_PN} alongside the real part of the $h_{2,2}$ and
  $h_{2,1}$, and examine the learned coefficients of the $\vdot$
correction in Fig.~\ref{fig:NN_orbitalcorrection}.

This experiment establishes a baseline for the framework, demonstrating that
the neural network can successfully reproduce analytically known higher-order PN terms
before being extended to the full PN$\rightarrow$NR case described in Sec.~\ref{sec:pntonr}.



\section{PN \texorpdfstring{$\rightarrow$}{to} NR}
\label{sec:pntonr}
\subsection{Overview}
\label{subsec:overview}

Building upon the 2PN$\rightarrow$3PN experiment, we extend
the framework to bridge the gap between PN
approximations and NR waveforms. For the PN baseline, we adopt the
TaylorT4 approximant~\cite{Boyle:2007ft} truncated at 4.5PN order~\cite{Blanchet:2023bwj,Blanchet:2023sbv} as
implemented in the \texttt{PostNewtonian.jl}
package~\cite{Boyle_PostNewtonian_jl_2024}.

For the target NR data, we use the \NRSurOld~\cite{Varma:2018mmi} surrogate model, which
reproduces NR waveforms from the SXS catalog~\cite{Scheel:2025jct} with high fidelity and
serves as a computationally efficient proxy for direct NR simulations.
To construct the training dataset, we generate a set of nonspinning BBH
waveforms using \NRSurOld.
Specifically, we sample ten systems uniformly in symmetric mass ratio $\nu\in [0.109, 0.25]$,
corresponding to mass ratio $q\in[1,8]$.
Each waveform is initialized at a reference $(2,2)$ mode frequency of
$f_{2,2}=0.00497 [1/M]$, which approximately corresponds to an orbital
velocity $v\simeq 0.25$ in the PN expansion.

The dataset is then divided into eight training waveforms and two validation waveforms.\footnote{
  This small set of validation waveforms is a holdover from previous
  experiments and is not optimized for this problem.  Future work could
  explore more robust validation strategies.
}
These correspond to training mass ratios $q\in\{1.0, 1.7, 2.2, 2.6,  4.5, 5.4, 6.5, 8.0\}$ and
validation mass ratios $q\in\{3.2, 3.8\}$. The
training waveforms are used to compute the loss that drives optimization, while the validation
waveforms provide an independent validation loss that is used solely for monitoring generalization during
training. For the final assessment of model performance, we additionally generate a set of 100
waveforms uniformly sampled in $q \in [1,8]$ and evaluate the trained model on these waveforms. This
larger resampled set provides a more robust estimate of the final modeling error and of the model's
generalization quality across the mass ratio range.

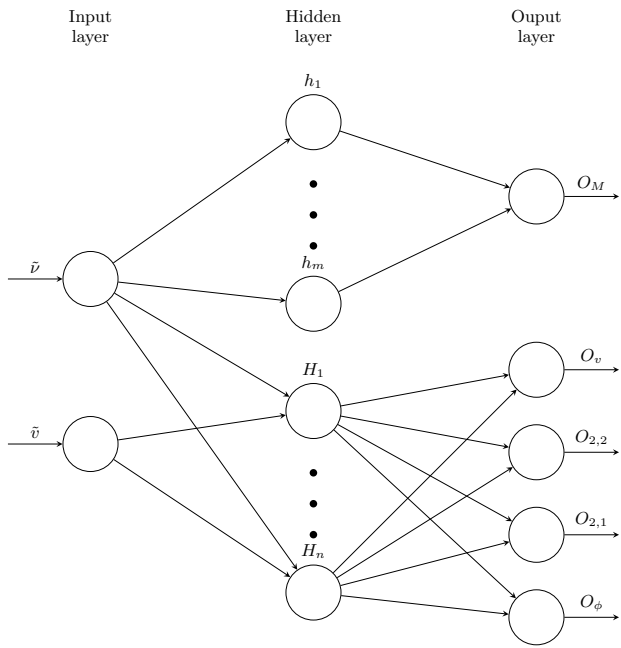
\begin{figure}[t]
\label{fig:pnnr_nn}
\centering
\makebox[\columnwidth][c]{%
  \resizebox{0.95\columnwidth}{!}{%
    {}
    \input{fig_pnnr_tikz.tex}%
  }%
}
\caption{Neural network architecture for the PN$\rightarrow$NR experiment.
The two inputs $(\nu,v)$ are linearly scaled to the interval
$[-1,1]$ to improve numerical stability
during training. Each output is further rescaled by a
tunable scaling coefficient before being applied as a correction to
the orbital dynamics or waveform modes. The final model
uses 4 neurons in the mass-correction branch and 12 neurons
in the orbital/mode-correction branch, totaling 101 trainable parameters.}
\end{figure}

Although the overall framework remains largely unchanged from the
2PN$\rightarrow
$3PN problem, one vital addition is a small neural network branch
that outputs $\nu$-dependent corrections to the total mass of the system, $M$.
The motivation for introducing this term is that the definition
of black hole masses does not translate directly between PN
and NR. In NR simulations, the black hole mass can be defined
as the Christodoulou mass computed from the apparent horizons at a reference time,
whereas in the PN approximation each black hole is treated as a point particle.
To account for discrepancies arising from inconsistent mass definitions,
we introduce a learned correction to the total mass.
The complete architecture of the extended neural network, including
this additional mass correction branch, is
shown in Fig.~\ref{fig:pnnr_nn}.

The overall PN$\rightarrow$NR procedure can be summarized as follows:
\begin{enumerate}
\item{Apply the total mass correction $O_M$ to $M_1,M_2$:\\
\begin{align}
\begin{aligned}
M_1 &= \frac{q}{1+q} \big(1.0 + \beta_{M} O_M(\nu)\big)\\
M_2 &= \frac{1}{1+q} \big(1.0 + \beta_{M} O_M(\nu)\big)
\end{aligned}
\end{align}
}
\item{Apply the orbital correction $O_v(\nu,v)$ to the 4.5PN expression for $\vdot$}
\item{Integrate the ODE with the corrected $\vdot$ to obtain the corrected orbital evolution.}
\item{Compute the waveform modes $h_{2,2}$ and $h_{2,1}$ from the corrected orbital solution.}
\item{Apply the waveform mode corrections as described in
Eq.~\eqref{eqn:waveform_correction}}
\end{enumerate}

As in the previous experiment, the inputs are linearly scaled
to the interval $[-1,1]$ to ensure well-conditioned
gradients during training. Each neural network output is further
multiplied by a tunable scaling coefficient $\beta$
before being applied as corrections to $M, \vdot,$ or the
waveform modes $h_{2,2}$ and $h_{2,1}$.
The specific values adopted in the final model are summarized in
Table~\ref{tab:hyperparams}. We note that these values
were empirically chosen on achieving stable convergence during training
and we did not perform a systematic hyperparameter optimization.
\begin{table}[tbp!]
\centering
\caption{Hyperparameters used in the final neural network model.}
\label{tab:hyperparams}
\begin{tabular}{lc}
\hline\hline
\textbf{Hyperparameter} & \textbf{Value} \\
\hline
$\beta_M$       &$10^{-3}$ \\[-3pt]
$\beta_v$       &$10^{2}$\\[-3pt]
$\beta_{2,2}$    &$10^{-2}$\\[-3pt]
$\beta_{2,1}$    &$10^{-2}$\\[-3pt]
$\beta_{\phi}$  &$10^{-3}$\\
\hline\hline
\end{tabular}
\end{table}
Several modifications to the loss function are also introduced
and are discussed in detail in Sec.~\ref{subsec:loss_functions}.

For the training, we initially tested a
two-stage strategy where we first trained on segmented waveform data
and then fine-tuned with full waveform data, similar to the 2PN$\rightarrow$3PN
problem in Sec.~\ref{sec:2pn3pn}. However, this method
proved ineffective for the present problem.\footnote{
Unlike the smooth and analytically defined
2PN$\rightarrow$3PN correction, the
PN$\rightarrow$NR discrepancy is highly nonlinear and
cumulative. Segment-wise training thus overemphasizes
local residuals and produces inconsistent corrections across
segments.}
Instead, introducing a new $v$-dependent
loss term yields more robust and efficient training, as we will discuss
in more detail in Sec.~\ref{subsec:loss_functions}.

\begin{figure}[htbp!]
  \centering
  \includegraphics[width=1.0\linewidth]{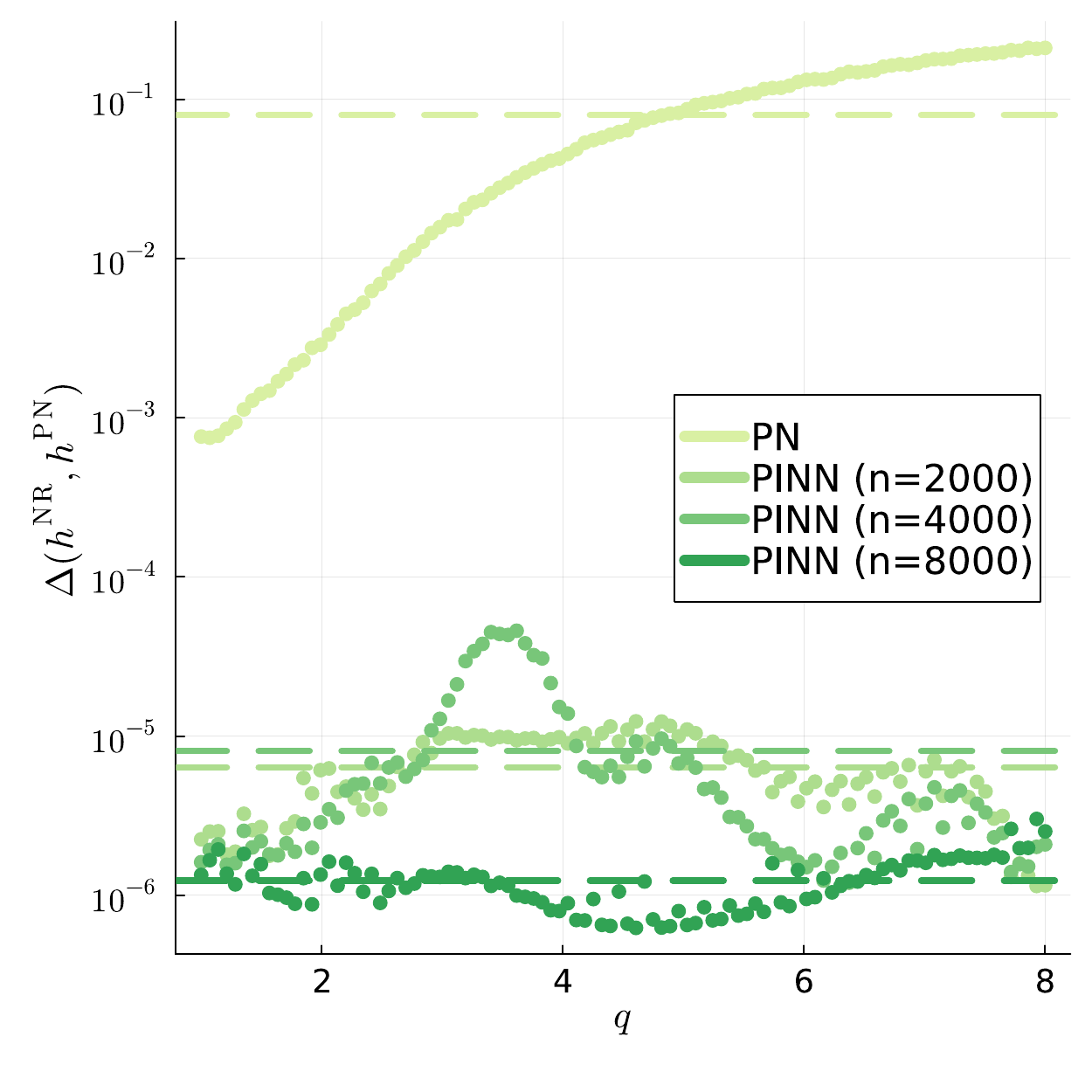}
  \caption{Evolution of the waveform mismatch during training for the
  PN$\rightarrow$NR model. Dashed lines indicate the average
  mismatch over all mass ratios $q$. The mismatch decreased rapidly during
  the early phase of training and eventually converges to a similar level across
  all values of mass ratio, $q$. This suggests that the network learns a
  consistent correction rather than overfitting to a specific mass ratio.}
  \label{fig:mismatch_improvement}
\end{figure}

\begin{figure}[!h]
  \centering
  \includegraphics[width=1.0\linewidth]{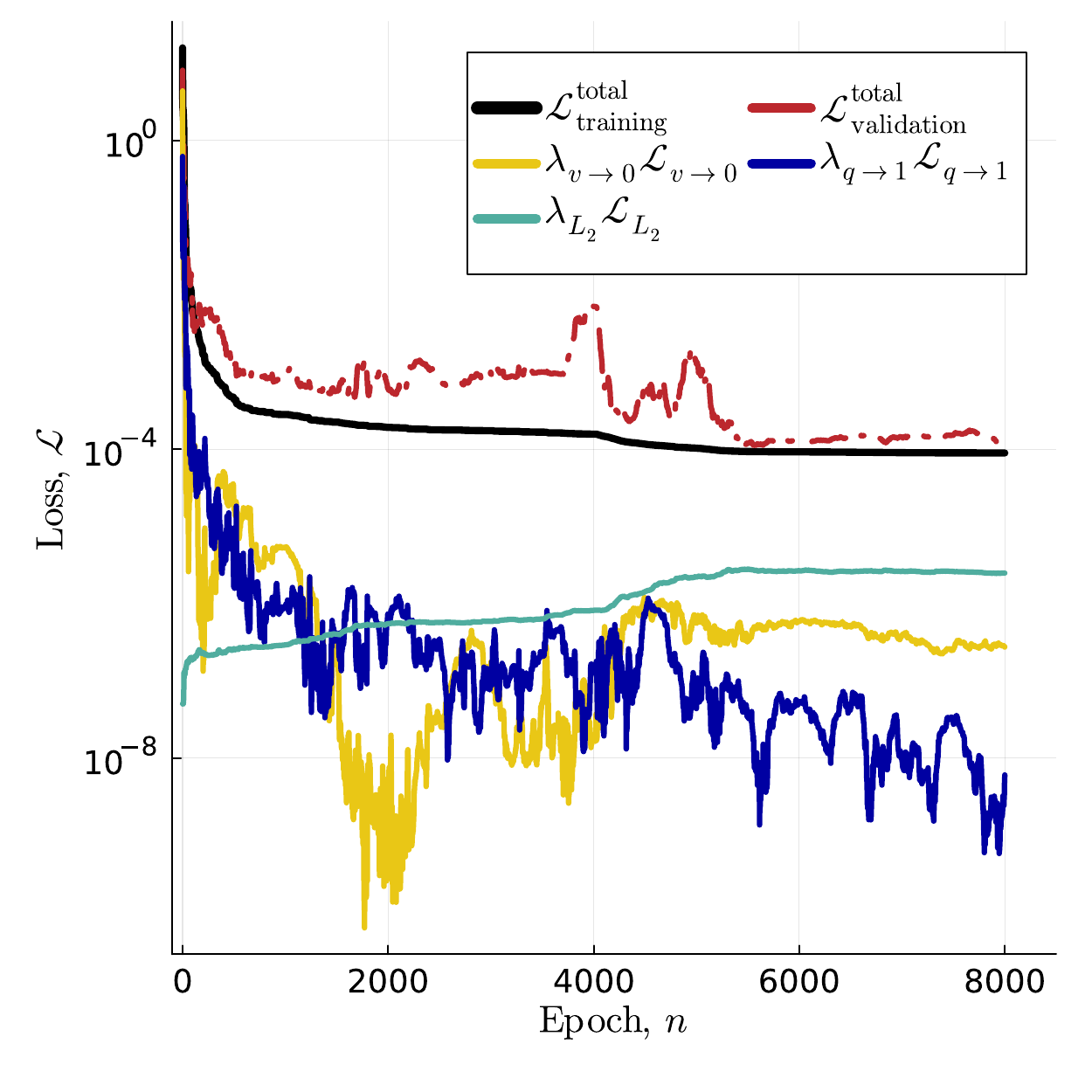}
  \caption{The history of the total loss and its main components. The losses decrease rapidly during the
  first 2000 epochs, after which a divergence between training and validation becomes apparent and peaks
  near 4000 epochs. A warm restart at this point resets the Hessian approximation, restores alignment
  between the two losses, and leads to improved generalization in the subsequent optimization.}
  \label{fig:loss_history}
\end{figure}

As mentioned earlier, we estimate the modeling error
by evaluating the network on a set of $100$ surrogate-generated waveforms
uniformly sampled in mass ratio $q \in [1,8]$. The evolution of these waveform
mismatches over training epochs is shown in Fig.~\ref{fig:mismatch_improvement}.
The model quickly reproduces fiducial surrogate waveforms with a mean
mismatch of order $10^{-5}$ after roughly $2000$ epochs. Beyond this point,
however, the training and validation losses begin to diverge, and the generalization
quality degrades. This discrepancy reaches a clear peak at epoch $4000$,
as shown in Fig.~\ref{fig:loss_history}, before beginning to decrease again with a
rapid drop in the validation loss. By epoch $8000$, the training and validation losses have
converged, and the resulting mismatches settle to a uniform spread at the
$10^{-6}$ level across the resampled waveforms.

The $4000$-epoch point coincides with a checkpoint
at which we restart the optimization. The warm restart retains the optimized parameters but
resets the inverse Hessian approximation and gradient information. To verify that the turnover
at $4000$ epoch is not an artifact of restarting, we additionally performed a continuous $8000$-epoch
run with no restarts. This run exhibits the same peak and subsequent decline at $4000$ epochs, confirming
that the feature is intrinsic to the optimization landscape, rather than the restart.
Notably, the checkpoint warm restart run shows better
agreement between the training and validation losses by epoch $8000$, suggesting that the restart
helps stabilize the late-stage optimization.

All model training was performed with an
AMD EPYC 9654 96-core processor. The implementation parallelizes the
workload over training samples. For the final model, which used eight training
waveforms, we employed 8 CPU cores. Using BFGS optimization, each epoch required
approximately 6.6 seconds of wall clock time, and the final model was trained
for 8000 epochs, corresponding to about 14.7 hours of wall clock time (120 CPU hours).

\subsection{Loss function}
\label{subsec:loss_functions}

The definition of the loss function plays a central role in ensuring that the
learned corrections remain physically meaningful and numerically stable across
the entire training domain. Beyond the standard mean-squared error between the
target and predicted waveforms, a couple additional physics-informed and regularization
components are incorporated.

We first describe how the waveform mismatches are
actually calculated and incorporated into the overall loss. The waveform
mismatch component of the total loss is defined as
\begin{equation}
\label{eqn:waveform_loss}
\lie_{\textrm{w} } = \bigg\langle
\frac{\lVert h_{2,2}^{\textrm{NR}} - h_{2,2}^{\textrm{PINN}}\rVert^2 +  \lVert h_{2,1}^
{\textrm{NR}} - h_{2,1}^{\textrm{PINN}}\rVert^2}
{\lVert h_{2,2}^{\textrm{NR}} \rVert^2+\lVert h_{2,1}^{\textrm{NR}} \rVert^2} \bigg\rangle,
\end{equation}
where $\langle \, \rangle$ denotes the average over all training samples and
$\Vert \, \rVert$ follows Eq.~\eqref{eqn:l2}.
Note that this form of waveform mismatch where the contributions
from $(2,2)$ and $(2,1)$ modes are summed represents
averaging waveform mismatch across all points on a sphere.

We next impose physically motivated regularization terms
that enforce the corrections to vanish in the appropriate limits.
Specifically, the orbital and waveform corrections
$O_v,\, O_{2,2}, \,  O_{2,1}, \, O_{\phi} \, \!\to\!0$ as $v\!\to\!0$ are required
to approach zero as $v\!\to\!0$ and the subdominant $(2,1)$
mode correction $O_{2,1}$ must also vanish as the system
approaches equal masses ($q\!\to\!1$).
To encode these behaviors, we introduce two additional loss terms,
$\lie_{v\to0}$ and $\lie_{q\to1}$, defined below.

The first term, $\lie_{v\to0}$ penalizes nonzero network outputs at $v=0$
across ten uniformly spaced mass ratios $q\in[1,8]$
\begin{align}
\lie_{v\rightarrow0} = \sum_{\nu(q=8)}^{\nu(q=1)} \bigg( &|O_v(\nu,0)|^2 +
|O_{2,2}(\nu,0)|^2 +|O_{2,1}(\nu,0)|^2 \nonumber \\&+|O_\phi(\nu,0)|^2\bigg)
  \label{eqn:lphysv}
\end{align}

The second term, $\lie_{q\to1}$ enforces that the $(2,1)$ mode correction
smoothly vanishes in the equal-mass limit. For this term, we evaluate
$O_{2,1}$ at ten uniformly spaced $v$ within the training range $v\!\in\![0.24,0.38]$
and penalize any nonzero outputs
\begin{equation}
\label{eqn:lphysq}
\lie_{q\rightarrow1} =\sum_{v=0.24}^{0.38} |O_{2,1}(0.25,v)|^2
\end{equation}

To further encourage physical consistency between the PN and
NR orbital dynamics, we introduce an auxiliary term that compares the orbital
velocity of PN evolution with that inferred from the $(2,2)$ mode frequency
of the NR waveform. This term promotes better phase alignment between the PN
and NR waveforms.

We define the NR orbital velocity as
\begin{equation}
  \label{eq:1}
  v_{\text{NR}}=\omega_{\textrm{orb}}^{1/3}
  =(\pi f_{2,2})^{1/3}.
\end{equation}
This is a combination of two approximations: 1.~Kepler's Law and 2.~that the
orbital and $(2,2)$-mode frequency can be trivially related. With this, the
orbital velocity loss term is
\begin{equation}
  \lie_{v} = \bigg\langle \frac{\lVert v_{\textrm{NR}} - v_{\textrm{PN}} \rVert^2}{\lVert
    v_{\textrm{NR}} \rVert^2} \bigg\rangle\,
\end{equation}
where $\langle \, \rangle$ again denotes the average over all training samples and
$\Vert \, \rVert$ follows Eq.~\eqref{eqn:l2}.

Lastly, we add $L_2$ regularization terms, proportional to the squared $L_2$ norms
of neural network weights (we exclude the biases). This term mitigates overfitting and
improves generalization. We can fine-tune the strength of this regularization term
with the hyperparameter $\lambda_{L_2}$.

Combining all terms, the total loss function is
\begin{align}
  \lie &=\lambda_w  \lie_w  +
          \lambda_{v}  \lie_{v}  +
          \lambda_{q\rightarrow1} \lie_{q\rightarrow1}  \nonumber \\
       &+  \lambda_{v\rightarrow0} \lie_{v\rightarrow0} +
         \lambda_{L_2} \lie_{L_2}\,.
\end{align}

The inclusion of these physically motivated components
plays a crucial role in constraining the behavior of learned corrections.
The vanishing-limit penalty for $v\rightarrow 0$ ensures that the network
respects known agreement between PN and NR in the early inspiral
and prevents unphysical extrapolations at small $v$ that are not
represented in the training data. Similarly, enforcing, the equal-mass limit for
the $(2,1)$-mode amplitude corrections ensures smooth and consistent behavior
as $q \rightarrow 1$, compensating for the fact that the physical $(2,1)$ mode
itself vanishes in this limit and thus lacks a natural boundary condition.

In the final model, we adopt the loss hyperparameter values listed in
Table~\ref{tab:loss_hyperparams}. These coefficients control the
relative weighting of the loss components and were selected through
empirical experimentation to achieve a balance between
physical fidelity, numerical stability, and optimization efficiency.
The relative magnitudes were tuned to ensure that the loss terms
and their gradients are of comparable scale at the start of training.
Small adjustments to these parameters can noticeably affect the
convergence behavior, reflecting the commonly observed sensitivity
of neural-network optimization to hyperparameter choices.

\begin{table}[h!]
\centering
\caption{Loss-function hyperparameters used in the final model.}
\label{tab:loss_hyperparams}
\begin{tabular}{lc}
\hline\hline
\textbf{Hyperparameter} & \textbf{Value} \\
\hline
$\lambda_w$        & $10^{2}$  \\[-3pt]
$\lambda_v$        & $10^{3}$  \\[-3pt]
$\lambda_{v\to0}$  & $1$              \\[-3pt]
$\lambda_{q\to1}$  & $1$              \\[-3pt]
$\lambda_{L_2}$    & $10^{-9}$ \\
\hline\hline
\end{tabular}
\end{table}

\subsection{Extrapolation}
\label{subsec:extrapolation}

\begin{figure}[h!]
\label{fig:extrap_in_q}
\includegraphics[width=0.5\textwidth]{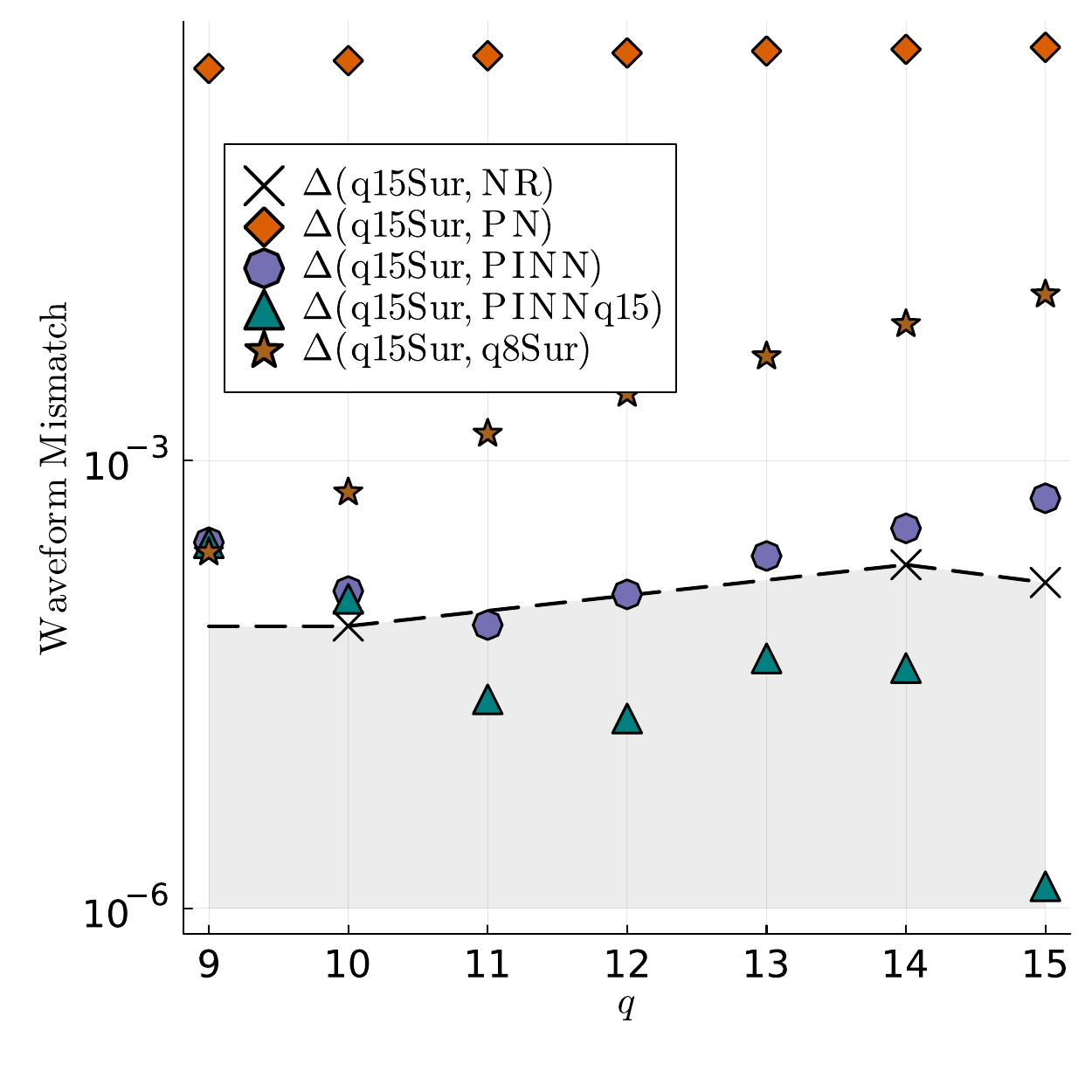}
\caption{Waveform mismatches for mass ratios $q\geq 8$ computed against the \texttt{NRHybSur2dq15},
surrogate. Results are shown for the PN baseline, PINN model trained on $q\in[1,8]$,
its extended version with an additional training point at $q=15$, and the
\texttt{NRHybSur3dq8} surrogate. Although both the original PINN and
\texttt{NRHybSur3dq8} were trained over the same range, the PINN model
exhibits better agreement with \texttt{NRHybSur2dq15} at high mass ratios.
Incorporating a single additional $q=15$ training point further improves waveform
accuracy across the intermediate region ($9 \lesssim q \lesssim 15$),
reducing mismatches toward the reference level.
The dashed line denotes the mismatch level between NR and \texttt{NRHybSur2dq15}, which
 serves as the effective reference limit for extrapolation accuracy.
 }
 \end{figure}

In addition to modeling well within the training region, the model exhibits encouraging
signs of extrapolation capabilities. As shown in Fig.~\ref{fig:extrap_in_q}, the
trained NN corrections significantly reduce the mismatch between PN and NR
waveforms at higher mass ratios ($q \geq 8$), improving from the $10^{-1}$ level
to roughly $10^{-4}$--$10^{-3}$. Even though the NN was trained only
on the same range ($q\in[1,8]$) as \NRSurOld, it performs
noticeably better in this higher $q$ regime.

We use the \texttt{NRHybSur2dq15} surrogate~\cite{Yoo:2022erv}, which is trained on hybrid PN-NR
waveforms up to mass ratio $q=15$, as a proxy for NR to assess extrapolation performance.
For reference, we also include the mismatch between NR simulations and
\texttt{NRHybSur2dq15}. Since the \texttt{NRHybSur2dq15} serves here as the effective
``ground truth,'' it would not be meaningful
for our PINN model to achieve errors below this reference. The shaded
region therefore marks the range below which further improvement has no physical significance.
While our results extend slightly outside of this region, their close proximity demonstrates that
the framework generalizes reasonably well beyond its training region.

\subsection{Extending the model}
\label{subsec:extension}

Building on the model's demonstrated extrapolation capability, we
next investigate how incorporating an additional training data affects its
performance. In the original setup, the network was trained on eight systems
with mass ratios $q\in [1,8]$. We now introduce a single
additional training waveform at $q=15$, well separated from the original training range, and examine how
the model's accuracy changes across intermediate mass ratios.

To isolate the impact of this new data point, we keep the original
training dataset unchanged and simply append the $q=15$ waveform
generated using \texttt{NRHybSur2dq15}. Starting from the previously trained network
parameters, we resume optimization for an additional 4000 epochs using the same loss function
and hyperparameters. This warm start allows the network to incorporate new high-$q$ constraints
while preserving the structure learned over $q\in [1,8]$. Throughout this extended training, we verify that
the model's performance within the original training region remains stable.

As shown by the green triangles in Fig.~\ref{fig:extrap_in_q}, incorporating a
single additional training waveform
greatly improves the model's agreement with \texttt{NRHybSur2dq15}
for $q\in[9,14]$. This result demonstrates the framework's
potential to be efficiently extended by incorporating only a small number of sparsely
sampled training data, enabling systematic improvement of model accuracy
without requiring dense coverage of the parameter space.

\section{Conclusion}
\label{sec:conclusion}

We developed a physics-informed neural network by adding a neural network to the
post-Newtonian (PN) orbital evolution equations and waveform computations to
reduce the difference between PN and numerical relativity (NR). Using the 4.5PN
TaylorT4 PN waveforms as the baseline PN model and restricting to nonspinning
systems, we are able to reduce the mismatch between PN and NR from $0.1$ to
$1\times10^{-6}$. Our loss function not only includes the waveform mismatch, but
also several terms that weakly enforce physical constraints to aid in extrapolation,
as well as an $L_2$ regularization term of the NN weights.

A notable aspect of this study is that the model achieves these results while
being trained on only eight waveforms. This demonstrates
the data efficiency made possible by embedding physical structure directly into the
model, and suggests that such a framework may enable the direct use of NR
simulations for training even when only a sparse set of waveforms is available.
Since the model we present here is physics-informed, i.e., it solves PN
equations with (small) corrections, one would expect improved extrapolation
compared to a more data-driven model like surrogates. We find that the
network maintains lower mismatch than the surrogate model (\NRSurOld)
outside of the training region, achieving $10^{-4}-10^{-3}$ mismatch levels.
Moreover, adding a single additional training point at $q=15$ substantially
improves performance across intermediate mass ratios, demonstrating that
the framework can be efficiently extended with only sparsely sampled data.

Our results demonstrate that physics-informed neural networks provide a
promising avenue for bridging the gap between PN theory and NR. Future work
will focus on extending the current framework to include spin and
eccentricity, as well as applying it through merger and ringdown.

Software: 
  This project made use of \texttt{Julia} packages including
  \texttt{PostNewtonian.jl}~\cite{Boyle_PostNewtonian_jl_2024}
  \texttt{Lux.jl}~\cite{pal2023lux},
  \texttt{Optimization.jl}~\cite{vaibhav_kumar_dixit_2023_7738525},
  \texttt{ForwardDiff.jl}~\cite{RevelsLubinPapamarkou2016},
  \texttt{DifferentialEquations.jl}~\cite{diffeqnjl}, and
  \texttt{Plots.jl}~\cite{plotjl}.
  This project made use of \texttt{Python} packages including
  \texttt{gwsurrogate}~\cite{Field:2025isp},
  \texttt{NumPy}~\cite{harris2020array}, and
  \texttt{SciPy}~\cite{2020SciPy-NMeth}.

\begin{acknowledgments}
  This material is based upon work supported by the National Science Foundation
  under Grants No.~PHY-2407742, No.~PHY-2207342, and No.~OAC-2209655. Any
  opinions, findings, and conclusions or recommendations expressed in this
  material are those of the author(s) and do not necessarily reflect the views
  of the National Science Foundation. This work was supported by the Sherman
  Fairchild Foundation.

\end{acknowledgments}

\section*{DATA AVAILABILITY}

The data that support the findings of this article are openly available~\cite{zenodo_dataset}.

\bibliographystyle{apsrev4-2}
\bibliography{References}

\end{document}

%% file: fig_2pn3pn_tikz.tex
\begin{tikzpicture}[x=1.5cm, y=1.5cm, >=stealth]

\foreach \m/\l [count=\y] in {1,2}
  \node [neuron/.try, neuron \m/.try] (input-\m) at (0,1-\y) {};

\foreach \m [count=\y] in {1,missing,2}
  \node [neuron/.try, neuron \m/.try ] (hidden-\m) at (1.5,1.8-\y*1.25) {};
  
 \foreach \m [count=\y] in {1,2,3,4}
  \node [neuron/.try, neuron \m/.try ] (output-\m) at (3,2-\y) {};

\foreach \l [count=\i] in {1,n}
  \node [above] at (hidden-\i.north) {$H_\l$};
  
\foreach \l [count=\i] in {1,2}{
  \ifnum\i=1
    \draw [<-] (input-\i) -- ++(-1,0)
      node [above, midway] {$\tilde{\nu}$};
  \else
    \draw [<-] (input-\i) -- ++(-1,0)
      node [above, midway] {$\tilde{v}$};
  \fi
}
  
 \foreach \i in {1,2}
  \foreach \j in {1,...,2}
    \draw [->] (input-\i) -- (hidden-\j);
    
   \foreach \i in {1,...,2}
  \foreach \j in {1,2,3,4}
    \draw [->] (hidden-\i) -- (output-\j);
    
\foreach \i in {1,2,3,4}{
  \ifnum\i=1
    \draw [->] (output-\i) -- ++(1,0)
      node [above, midway] {$O_v$};
  \fi
  \ifnum\i=2
    \draw [->] (output-\i) -- ++(1,0)
      node [above, midway] {$O_{2,2}$};
  \fi
  \ifnum\i=3
    \draw [->] (output-\i) -- ++(1,0)
      node [above, midway] {$O_{2,1}$};
  \fi
  \ifnum\i=4
    \draw [->] (output-\i) -- ++(1,0)
      node [above, midway] {$O_{\phi}$};
  \fi
}
\foreach \l [count=\x from 0] in {Input, Hidden, Ouput}
  \node [align=center, above] at (\x*1.5,1.6) {\l \\ layer};

\end{tikzpicture}

%% file: fig_pnnr_tikz.tex
\begin{tikzpicture}[x=1.5cm, y=1.5cm, >=stealth]
\foreach \m/\l [count=\y] in {1,2}
  \node [neuron/.try, neuron \m/.try] (input-\m) at (0,1-2*\y) {};

\foreach \m [count=\y] in {1,missing,2}
  \node [neuron/.try, neuron \m/.try ] (ch_\m) at (2.7,2-\y*1.1) {};

\foreach \m [count=\y] in {1}
  \node [neuron/.try, neuron \m/.try ] (co_\m) at (5.4,1.25-\y*1.25) {};
  
 \foreach \i in {1}
  \foreach \j in {1,...,2}
    \draw [->] (input-\i) -- (ch_\j);
    
 \foreach \i in {1,2}
  \foreach \j in {1}
    \draw [->] (ch_\i) -- (co_\j);

\foreach \l [count=\i] in {1,2}{
  \ifnum\i=1
    \draw [<-] (input-\i) -- ++(-1,0)
      node [above, midway] {$\tilde{\nu}$};
  \else
    \draw [<-] (input-\i) -- ++(-1,0)
      node [above, midway] {$\tilde{v}$};
  \fi
}

\foreach \l [count=\i] in {1}{
    \draw [->] (co_\i) -- ++(1,0)
      node [above, midway] {$O_M$};
}

\foreach \l [count=\i] in {1,m}
  \node [above] at (ch_\i.north) {$h_\l$};

\foreach \m [count=\y] in {1,missing,2}
  \node [neuron/.try, neuron \m/.try ] (oh_\m) at (2.7,-1.5-\y*1.1) {};

\foreach \m [count=\y] in {1,2,3,4}
  \node [neuron/.try, neuron \m/.try ] (oo_\m) at (5.4,-1.1-\y*1.0) {};

\foreach \l [count=\i] in {1,2,3,4}{
  \ifnum\i=1
    \draw [->] (oo_\i) -- ++(1,0)
      node [above, midway] {$O_v$};
  \fi
  \ifnum\i=2
    \draw [->] (oo_\i) -- ++(1,0)
      node [above, midway] {$O_{2,2}$};
  \fi
  \ifnum\i=3
    \draw [->] (oo_\i) -- ++(1,0)
      node [above, midway] {$O_{2,1}$};
  \fi
  \ifnum\i=4
    \draw [->] (oo_\i) -- ++(1,0)
      node [above, midway] {$O_\phi$};
  \fi
}

  \foreach \l [count=\i] in {1,n}
  \node [above] at (oh_\i.north) {$H_\l$};

 \foreach \i in {1,2}
  \foreach \j in {1,...,2}
    \draw [->] (input-\i) -- (oh_\j);

 \foreach \i in {1,2}
  \foreach \j in {1,2,3,4}
    \draw [->] (oh_\i) -- (oo_\j);

\foreach \l [count=\x from 0] in {Input, Hidden, Ouput}
  \node [align=center, above] at (\x*2.7,1.75) {\l \\ layer};

\end{tikzpicture}